\documentclass[12pt,preprint]{aastex}

\usepackage{natbib}
\newcommand{\Lya}{Ly$\alpha$\ }
\newcommand{\lya}{Ly$\alpha$\ }
 
\newcommand{\cm}{\ensuremath{{\rm cm}}} 
\newcommand{\erg}{\ensuremath{{\rm erg}}} 
\newcommand{\Hz}{\ensuremath{\, {\rm Hz}}} 
\newcommand{\sr}{\ensuremath{\, {\rm sr}}} 
\newcommand{\msun}{\ensuremath{{M_\sun}}}

\shorttitle{\Lya and 21cm}

\begin{document}

\title{The Spin-Kinetic Temperature Coupling and the Heating Rate due to
Lyman Alpha Scattering before Reionization: Predictions
for 21cm Emission and Absorption}


\author{Xuelei Chen}
\affil{The Kavli Institute for Theoretical Physics, UCSB, Santa
Barbara, CA 93106, USA}
\email{xuelei@kitp.ucsb.edu}
\and
\author{Jordi Miralda-Escud\'e}
\affil{Department of Astronomy, The Ohio State University,
Columbus, OH 43210, USA }
\email{jordi@astronomy.ohio-state.edu}

\begin{abstract}

  We investigate the interaction of \Lya photons produced by the first
stars in the universe with intergalactic hydrogen prior to reionization.
The background \Lya spectral profile is obtained by solving a
Fokker-Planck equation. Accurate values of the heating and scattering
rates, and the spin-kinetic temperature coupling coefficient, are
presented. We show that the heating rate induced by the \Lya scatterings
is much lower than found previously, and is basically negligible. The
dominant heating source is most likely the X-rays from the first
ionizing sources, which are able to penetrate into the atomic medium.
The scattering of \Lya photons couples the hydrogen spin temperature to
the kinetic temperature. If the first ionizing sources in the universe
did not emit significant X-rays, the spin temperature would be rapidly
brought down to the very low gas kinetic temperature, and a 21cm
absorption signal against the CMB larger than 100 mK would be predicted.
However, we argue that sufficient X-rays are likely to have been emitted
by the first stellar population, implying that the gas kinetic
temperature should rapidly increase, turning a reduced and brief
absorption signal into emission, with a smaller amplitude of $\sim 10$
mK. The detection of the 21cm absorption and emission feature would be a
hallmark in unravelling the history of the ``dark age'' before
reionization.

\end{abstract}


\section{Introduction}

  After recombination of the primordial plasma at $z \sim 1000$, 
the Universe entered an age of darkness. The primordial density
fluctuations continued to grow, until non-linear gravitational collapse
led to the formation of the first dark matter halos in which the gas
could collapse, and then radiatively cool. The dark ages ended at the
time when the first stars formed in these halos, with a characteristic
mass of $\sim 10^6 \msun$ \citep{CR86,T97,abn00,bcl02}. These first
stars, as well as any black holes
or neutron stars formed subsequently which could accrete surrounding
matter, produced photons capable of ionizing hydrogen (with wavelength
$\lambda < 91.2 \, {\rm nm}$). As soon as the ionizing radiation
produced in the sites of star formation was able to escape into the
low-density intergalactic medium, the process of reionization started.
During reionization, the ionized volume fraction of the universe
increased gradually, until all the low-density regions were ionized.
Recent observations of the Gunn-Peterson trough
\citep{GP65} at $z\gtrsim 6$ \citep{b01,f02,f03}, indicate that the end
of reionization probably occurred at $z\simeq 6$, although the epoch at
which reionization began is still unknown (e.g., \citealp{G00,M02}).

  The redshifted 21 cm line opens up a promising window for observing
this period of cosmic history. This line is produced by the transition
between the singlet and triplet hyperfine levels of the hydrogen atom at
the electronic ground state. The line can be observed in absorption or
emission against the Cosmic Microwave Background (CMB). The brightness
temperature is given by
\begin{equation}
\delta T = {T_s - T_{CMB} \over 1+z}\, (1-e^{-\tau}) \simeq
(0.025 \, {\rm K})\, \left( {\Omega_b h_0\over 0.03} \right)\,
\left( {0.3\over \Omega_{m0}} \right)^{1/2}
\left( {1+z\over 10} \right)^{1/2}\, {\rho_{HI}\over \bar\rho_H}\,
{ T_s - T_{CMB} \over T_s} ~.
\end{equation}
where $T_s$ and $T_{CMB}$ are the hydrogen spin temperature and the
CMB temperature, $\tau$ is the optical depth, $\rho_{HI}$ is the
density of atomic hydrogen and $\bar\rho_H$ is the mean hydrogen
density of the universe (we have used the approximations $\tau \ll 1$
and $\Omega_{m0}(1+z)^3 \gg 1$). Therefore, the detailed
spectrum of the CMB at the long wavelengths of the redshifted 21cm,
when observed with high angular resolution, contains precious
information on the patchwork of density, spin temperature, and
ionization inhomogeneities during the era when the intergalactic
medium was still atomic \citep{HR79,SR90,KPS95,MMR97,SWMdB99,TMMR00,ISFM02}.

  The hydrogen spin temperature plays a key role in determining the
amplitude of the 21cm emission or absorption signal against the CMB.
At the same time, the evolution of the spin temperature depends on the
gas kinetic temperature, and the way that the hyperfine structure level
populations couple to the CMB and the kinetic temperatures. The adiabatic
cooling of the hydrogen due to the expansion of the universe causes the
gas temperature to decline as $(1+z)^2$, while the CMB temperature is
decreasing only as $1+z$, so the gas cools below the CMB. At redshifts
$z\gtrsim 100$, the residual ionization left over from the epoch of
recombination \citep{P68} keeps the gas kinetic temperature at the
CMB temperature owing to the heating produced by electron scattering
of the CMB, but below this redshift this heating source becomes small
and the gas temperature drops below the CMB. If the spin temperature can
be coupled to the kinetic temperature at this stage, the atomic medium
would be observed in absorption.

  The coupling of the spin and kinetic temperatures can be achieved by
atomic collisions or by scattering of \lya photons. The coupling due to
atomic collisions is small except at very high redshift, 
where observations of the redshifted 21cm radiation becomes more difficult,
or for unrealistically high values of $\Omega_b$ \citep{SR90}. A
stronger coupling can be induced by \lya photons, which should be
produced as soon as the first stars are formed. The \lya photons couple
the spin and kinetic temperatures by being repeatedly scattered in the
gas. The resonance scattering of a photon consists of a transition from
$n=1$ to $n=2$, followed by the opposite downward transition. This
scattering process can change the population of the hyperfine structure
levels \citep{W52,F58}. The probabilities for a transition back to the
ground or excited hyperfine structure states depend on the slope of the
radiation spectrum near the \lya line center, or the ``color
temperature''. When the radiation spectrum reaches a steady state, this
color temperature near the \lya line center is equal to the gas kinetic
temperature, so the \lya scatterings introduce a thermal coupling
between the spin and kinetic temperature. The result of these couplings
is that the spin temperature is given by \citep{F58,F59}
\begin{equation}
\label{eq:Ts}
T_S = \frac{T_{CMB}+y_{\alpha} T_k + y_c T_k}{1+y_{\alpha}+ y_c} ~,
\end{equation}
where
\begin{equation}
\label{eq:y}
y_{\alpha} \equiv \frac{P_{10} T_*}{A_{10} T_k}; \qquad 
y_c \equiv \frac{C_{10} T_*}{A_{10} T_k} ~.
\end{equation}
Here, $T_k$ is the kinetic temperature, $T_* = 0.0682$K is the hyperfine
energy splitting, $A_{10} = 2.87 \times 10^{-15} \, {\rm s}^{-1}$ is the
spontaneous emission coefficient of the 21cm line, and
$C_{10}=\kappa_{10} n_H$ is the
collisional de-excitation rate of the excited hyperfine level. The
value of $\kappa_{10}$ ranges from $2\times 10^{-14} $
to $2.5\times 10^{-10} \cm^3\, {\rm s}^{-1}$ for $T$ in the range  
$1-1000$ K and was tabulated in \citet{AD69}. 
 
The indirect de-excitation rate
$P_{10}$ of the hyperfine structure levels is related to the total \Lya
scattering rate $P_{\alpha}$ by $P_{10}=4 P_{\alpha}/27$ \citep{F58}, 
which is given by
$P_{\alpha}=\int d\nu c\, n_{\nu}\,\sigma(\nu)$,
where $n_{\nu}$ is the number density of photons per unit frequency,
and $\sigma(\nu)$ is the cross section for \lya scattering
(\citealp{MMR97}, hereafter MMR).

  Besides coupling the spin temperature and kinetic temperature of the
gas, the scattering of \Lya photons may also heat up the gas. In this
paper, we present the scattering rate and heating rate due to \lya
photons in a uniform atomic intergalactic medium as a function of the
two parameters they depend on: the gas temperature and the \lya
scattering optical depth of the medium (or Gunn-Peterson optical depth).

  There are two different ways in which photons propagating through the
atomic medium may reach the \lya line, and start being repeatedly
scattered by hydrogen: (a) Photons that are emitted between the \lya and
Ly$\beta$ wavelengths by sources such as massive stars will be
redshifted as they penetrate the atomic medium until they reach the \lya
wavelength. These photons enter the \lya line from its blue wing, and
will be called {\it continuum} photons in this paper. (b) Photons that
are emitted at wavelengths shorter than Ly$\beta$, but still longer than
the Lyman limit so that they can penetrate deep into the atomic medium,
will redshift until they reach the Ly$\beta$ or a higher Lyman series
line, and will then be converted into a Ly$\alpha$ photon when the
hydrogen atom they excite decays first to the 2p state before reaching
the ground state. These photons will be injected near the \lya line
center with a distribution determined by the Voigt function, and will be
called {\it injected} photons. These two types of photons result in
different scattering and heating rates, so results will be presented
separately for both of them.

  MMR estimated the heating rate due to \Lya photons by assuming that
the average relative change in a \Lya photon energy in each scattering
is the same as when the photon is scattered by an atom at rest:
$\langle \Delta E/E \rangle = -h\nu_{\alpha}/m_H c^2$. Hence, they
assumed that the total energy transfer rate is
\begin{equation}
\label{eq:MMRheatrate}
\dot{E} = -\langle \frac{\Delta E}{E} \rangle h \nu_{\alpha} P_{\alpha}.
\end{equation}
At the thermalization rate $P_{th} =27 A_{10} T_{CMB}/4T_*$ (which is
the rate $P_{\alpha}$ required to bring the spin temperature close to
the kinetic temperature), the heating rate assumed by MMR is
$\dot{E}/k_B \simeq 220 {\rm K\, Gyr}^{-1} (1+z)/7 $, which would heat
up the gas kinetic temperature above the CMB temperature in a fraction
of a Hubble time at $z\sim 6$. 

  In the present investigation we calculate the heating rate of the
continuum and injected \Lya photons by computing the background spectrum
near the \lya line using the Fokker-Planck method introduced by
\citet{RD94} (hereafter RD), and using energy conservation. We find that
the heating rate due to \Lya scattering is much smaller than the
estimate of eq.\ (\ref{eq:MMRheatrate}), and as a consequence \Lya
photons do not constitute a significant heating source of the gas.
Calculating the heating rate caused by \lya photons as if every atom
were at rest, without taking into account the thermal motions, is
incorrect, and as we shall see the heating rate per unit volume is not
proportional to the atomic density. We explain in the Appendix the
reason why the argument used by MMR in their Appendix B to justify using
eq.\ (\ref{eq:MMRheatrate}) for the heating rate is invalid. As pointed
out by MMR, X-rays are important in pre-heating the atomic medium before
it is reionized. In fact, our work leads to the conclusion that X-rays
are the only important source of heating before the atomic medium is
reached by an ionization front.
 
  This paper is organized as follows: in \S 2 we describe our method of
calculation. We present our result on scattering and heating rates in
\S 3, explaining the physical reason why gas is not heated by \lya
photons as if every atom were at rest. We illustrate the consequences
for future 21cm observations in \S 4, with an example for the evolution
of the spin temperature with redshift, including also a model for X-ray
heating. Our conclusions are summarized in \S 5. We adopt a flat
$\Lambda$CDM model with the following cosmological parameters:
$h_0=0.7$, $\Omega_{\Lambda}=0.7$, $\Omega_{m}=0.3$, $\Omega_b h_0^2=
0.021$, $Y_{\rm He}=0.24$.

The code for solving the line profile and calculating the
scattering and heating rates is available from the 
authors\footnote{Our code, named ``Lyman Alpha scattering and Spin
Temperature'', or {\tt LAST}, can be downloaded at 
\url{http://theory.kitp.ucsb.edu/$\sim$xuelei/LAST}.}.

\section{Method of Calculation}

  Our goal in this work is to compute the \lya scattering rate and the
heating rate of hydrogen induced by this scattering, from which one
can then compute the evolution of the spin temperature. The scattering
and heating rates can be calculated once the background spectrum near
the \lya line is known. Let us consider the background of photons
produced by constant, flat spectrum sources in an expanding universe (in
practice the intrinsic source spectrum does not matter because \lya
scattering is important only on a very narrow range of frequencies
around \lya, in which the intrinsic source spectrum can be considered to
be constant). In the absence of scattering, the photons would simply
redshift and would produce a photon number density per unit frequency
$n(\nu) = n_0$ that is independent of frequency near \lya. The presence
of scatterings by hydrogen introduces a feature in the spectrum around
the \lya line, which has a constant shape $n(\nu)$ once a steady-state
is reached. Given this spectrum, the rate at which \lya photons are
being scattered is given by the integral 
\begin{equation}
P_{\alpha} = c \int d\nu\, n(\nu)\sigma(\nu) ~.
\end{equation}
Here, $\sigma(\nu)$ is the \Lya scattering cross section as a function
of frequency. The cross section has a thermal width $\Delta \nu_D$
determined by the temperature $T$ of hydrogen atoms of mass $m_H$,
\begin{equation}
\Delta\nu_D=\frac{ \sqrt{2 k_B T/m_H} }{c}\, \nu_{\alpha} ~,
\end{equation}
where $\nu_{\alpha}$ is the \Lya line central frequency and $k_B$ is
the Boltzmann constant. The cross section at the line center is
\begin{equation}
\sigma_0 = \frac{\pi e^2}{m_e c} f_{12} [\Delta\nu_D]^{-1} ~,
\end{equation}
where $f_{12} = 0.416$ is the oscillator strength for the \Lya
transition. The dependence of the cross section on frequency is given
by the normalized Voigt function,
\begin{equation}
\phi(x) = \frac{a}{\pi^{3/2}} 
\int_{-\infty}^{\infty} d y \frac{e^{-y^2}}{(x-y)^2 +a^2} ~,
\end{equation}
where $x=(\nu-\nu_{\alpha})/\Delta\nu_D$, the Voigt parameter is
$a=A_{21}/(8\pi\Delta\nu_D)$, and
$A_{21} = 6.25 \times 10^{8}$ Hz is the Einstein spontaneous emission
coefficient.
We will also use the photon intensity $J(\nu) = c\, n(\nu)/(4\pi)$.
With these definitions, the scattering rate is
\begin{equation}
\label{eq:scatrate1}
P_{\alpha} = 4\pi\sigma_0\, \Delta\nu_D\, \int J(x) \phi(x) dx ~.
\end{equation}

  The heating rate of the gas induced by \lya scattering can be simply
obtained by requiring conservation of energy. The scatterings cause
the cosmic background to have a narrow spectral feature around the
\lya wavelength. Were the scatterings to cease at a certain instant,
this feature would simply be redshifted to longer wavelengths.
Instead, when a steady-state spectrum is reached, the feature is
fixed at a constant wavelength. The reduction in energy per unit volume
of the background due to this feature is $\Delta \epsilon =
\int d\nu [n_0 - n(\nu)]\, h\nu\, d\nu$ (for the purpose of this
discussion we ignore the photons injected at the \lya wavelength by
recombinations, so that in the absence of scatterings, $n(\nu) = n_0$;
injected photons will be fully considered later). Over an interval of
time $dt$, the energy per unit volume that the background photons must
lose to keep the spectral feature at a fixed wavelength is $\Delta
\epsilon (d\nu / \nu) = \Delta\epsilon H\, dt$, where $d\nu$ is the
change in frequency due to redshift over the interval $dt$, and $H$ is
the Hubble constant at time $t$. The energy lost by the photons must be
gained by the hydrogen atoms, so the gas heating rate per unit volume is
\begin{equation}
\label{eq:heatrate1}
\Gamma = H \Delta\epsilon = H\, \int \left[ n_0-n(\nu) \right]
\,h\nu\, d\nu = {4\pi H h\, \Delta\nu_D \over c}\, \int
\left[ J_0-J(\nu) \right]\,\nu\, d\nu ~.
\end{equation}
We use these equations in \S 3 to compute the scattering and heating
rates once the line spectrum has been calculated. An alternative
derivation of the heating rate in equation (\ref{eq:heatrate1}) is given
in the Appendix.

  The spectrum of the cosmic background near \lya is determined by the
radiative transfer equation for resonant scattering for a homogeneous
and isotropic expanding universe, which is (RD) 
\begin{equation}
\label{eq:rd}
\frac{\partial J(x,\tau)}{\partial \tau} = - \phi(x) J(x, \tau) 
+ \gamma_S \frac{\partial J(x,\tau)}{\partial x}
+ \int R_{II}(x,x') J(x',\tau) dx' 
 + C(\tau) \psi(x) ~,
\end{equation}
where $\tau= (c n_1 \sigma_0) t$ is the mean free time between
scatterings for a photon at the line center, $t$ is the physical time,
and $n_1\simeq n_H$ is the number density of hydrogen atoms in the
ground state. The Sobolev parameter is $\gamma_S = \tau_{GP}^{-1}$, where
$\tau_{GP} = (\pi e^2 n_1 f_{12})/(H m_e \nu_{\alpha})$ is the
Gunn-Peterson optical depth. The first term in the right-hand-side of
equation (\ref{eq:rd}) describes the removal of photons that are
scattered, and the second term accounts for the redshift. The third term
represents the addition of the photons redistributed over frequency
after scattering \citep{RH92}, and the fourth term is due to the
injection of new \Lya photons with a spectral profile $\psi(x)$. 

  In the Fokker-Planck approximation, the scattering problem is treated
as diffusion in the approximation that the width of the spectral
feature is large compared to the frequency change in typical
scatterings. The redistribution term then becomes (RD)
\begin{equation}
\int R_{II}(x,x') J(x',t) dx'=\phi(x) J(x)+\frac{1}{2}
\frac{\partial}{\partial x}\left[\phi(x)\frac{\partial J(x)}{\partial
x}+2\eta\phi(x) J(x) \right],
\end{equation}
where the recoil term is included, with $\eta = h\nu_{\alpha}/
( c \sqrt{2m_H k_BT} )$. This approximation conserves photon number.
The equation is then reduced to [see eq.\ (34) of RD]
\begin{equation}
\label{eq:FP}
\frac{\partial J}{\partial \tau}=\frac{1}{2} \frac{\partial}{\partial
x} \left[ \phi\frac{\partial J}{\partial x} + 2\eta \phi J\right]+
\gamma_S \frac{\partial J}{\partial x} + C(t) \psi(x).
\end{equation}
When the equilibrium state is reached, 
$\partial J/\partial \tau=0. $

  Since this equation is linear, we can write its solution as 
$J= J_c + J_i$, where $J_c$ is the spectrum of the continuum photons,
and $J_i$ is the spectrum of the recombination photons injected near
the line center.

  For the continuum photons, there is no injection and so $C(t)=0$.
Equation (\ref{eq:FP}) is reduced to
\begin{equation}
\phi(x) \frac{d}{d x} J_c(x)+ 2\,\left[\,\eta \phi(x)+\gamma_S \right]\,
J_c(x) = A,
\end{equation} 
where $A$ is a constant of integration determined by the boundary
condition: as $x \rightarrow \infty$, $\phi(x)\rightarrow 0$, 
$J_c \rightarrow J_{c0} $, where $J_{c0}= (4\pi/c)\, n_{c0}$ is the
unperturbed intensity. Therefore, $A=2\gamma_S J_0$. The equation is then
solved as an initial value problem of an ordinary differential equation
with the condition $J_c(x) \rightarrow J_{c0}$ for
$x \rightarrow - \infty$, required by photon number conservation.

\begin{figure}
\plotone{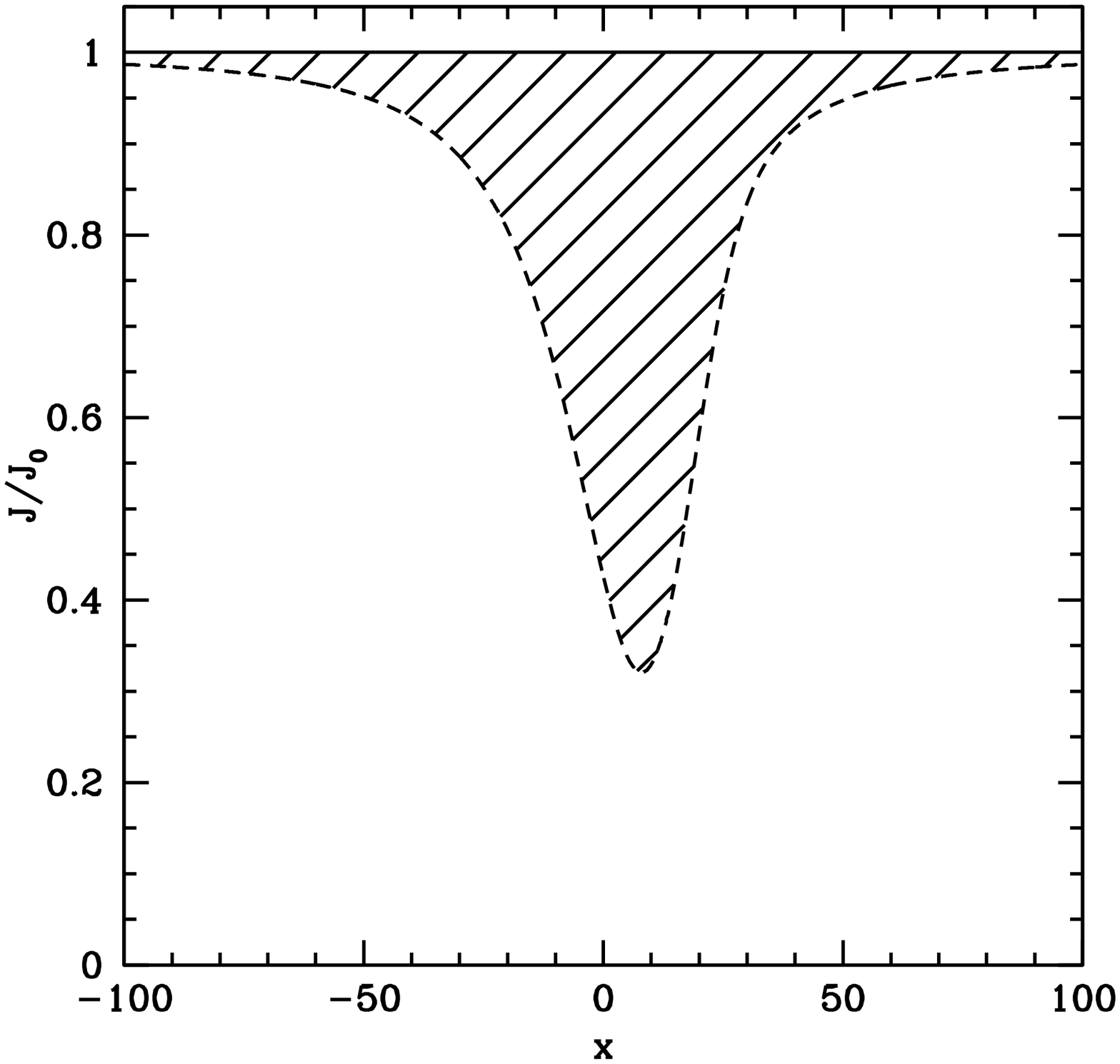}
\caption{\label{Jcon} Continuum \Lya photon spectrum. 
Solid line marks the flat (no-scattering) limit, 
dash line for spectrum produced with T=2.65 K, $\gamma_S =1.44 \times
10^{-6}$, corresponding to the condition of unheated gas
at $z=10$ in $\Lambda$CDM model.
The area of the shaded region gives the heating rate.
}
\end{figure}

  In Fig.~\ref{Jcon}, we plot the solution for a typical case:
$T=2.65$K and $\gamma_S =1.44 \times 10^{-6}$, corresponding to the
physical conditions of unheated gas at $z=10$ in a flat $\Lambda$CDM
model. An absorption feature is produced near the center of the line.

\begin{figure}
\plotone{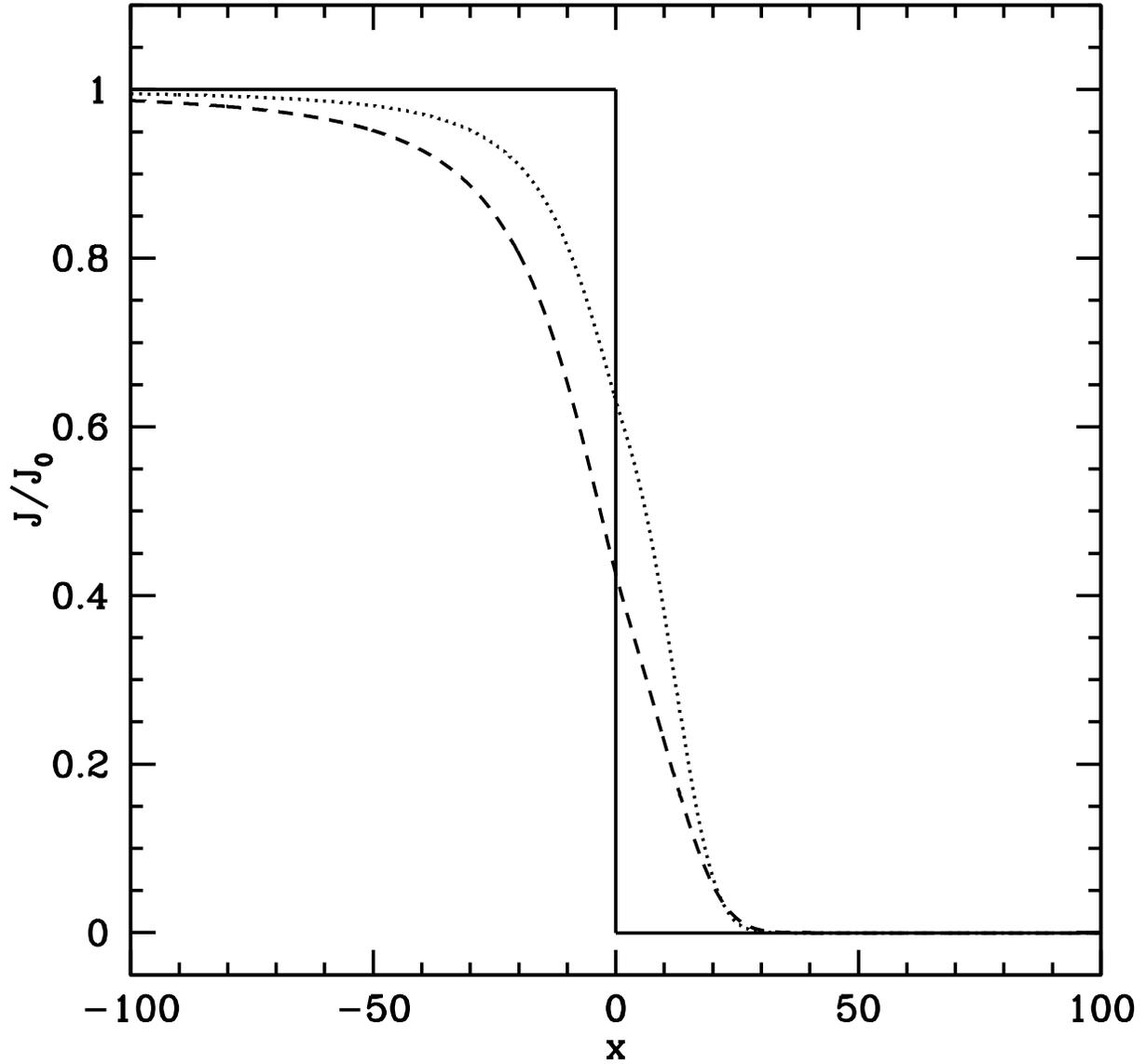}
\caption{\label{Jinj} Injected \Lya photon scattering spectrum. Solid
line shows the no-scattering limit. The dotted line for $T_k = 2.65$ K
gas at $z=10$ in which photon gains energy, 
dashed line for $T_k = 1$ K gas at $z=10$ in which photon looses energy.
}
\end{figure}

  For the injected photons, we assume a constant source $C(t) = C$.
This should generally be an accurate approximation because the timescale
for the evolution of the production rate of \lya photons from sources is
likely be of order the Hubble time, which is much longer than the
relaxation time, which is of order the time to redshift across the width
of the \lya
spectral profile of the cosmic background. We also inject photons with
a single frequency at the \lya line center, i.e., $\psi(x)=\delta(x)$.
Because photons change randomly in frequency over a thermal width at
each scattering, and the number of scatterings each photon undergoes
is very large, injecting the photons with a thermal width instead does
not make any difference to the final profile [we have verified this by
obtaining solutions with $\psi(x)$ given by a Voigt profile, which did
not result in any significant change to the resulting spectral profile].
The solution is 
\begin{equation}
\label{eq:injdelta}
\phi(x)\, \frac{d}{d x} J_i(x)+ 2\,\left[\,\eta \phi(x)+\gamma_S \right]
J_i(x) = -2 C \Theta(x) + B ~,
\end{equation} 
where $\Theta(x)$ is the Heaviside step function. The value of the
integration constant $B$ can again be determined from boundary
conditions: as $x \to +\infty$, we must now have $J_i(x) \to 0$, from
which we obtain $B = 2 C$. The profile of the line can again be obtained
by solving eq.\ (\ref{eq:injdelta}), with the initial condition
$J_i(x= -\infty) = J_{i0}$, where $J_{i0}$ is the photon intensity
due to the injected photons. The result is shown in Fig.~\ref{Jinj},
for two different temperatures; we
see that some of the photons injected at the line center are scattered
to the right (blue) side. 

  We note that the solution for the spectral profile of continuum and
injected photons has a characteristic width in the variable $x = \nu /
\Delta\nu_D$ that is substantially larger than unity, implying that the
Fokker-Planck approximation we have used should be accurate.

\section{Results}

  The heating and scattering rates are of course proportional to the
background intensity. During the epoch of reionization, at least one
ionizing photon has to be emitted for every baryon in the universe, and
more if the typical baryon can recombine many times. The number of \lya
photons emitted is of the same order for sources without a large Lyman
break, and can be much larger if many ionizing photons are internally
absorbed. We use the intensity corresponding to one photon per hydrogen
atom in the universe as a fiducial value:
\begin{equation}
\tilde J_0 (z) = \frac{n_H c}{4\pi \nu_{\alpha}} = 1.65\times 10^{-13}\,
(1+z)^3 \left(\frac{\Omega_b h_0^2}{0.02}\right) ~ \cm^{-2} {\rm s}^{-1} \sr^{-1} \Hz^{-1} ,
\end{equation}
where $n_H$ is the hydrogen number density.
For reference to other work, we quote also the value of the energy
intensity of this unit: 
$\tilde J_0\, h\nu_{\alpha} =
2.69\times 10^{-24}\, (1+z)^3 (\Omega_b h_0^2/0.02)\, \erg\, \cm^{-2}\,
{\rm s}^{-1} \sr^{-1} \Hz^{-1}$. The heating rate in equation
(\ref{eq:heatrate1}) can be
reexpressed as a heating rate per hydrogen atom and per Hubble time.
For the continuum photons, it is:
\begin{equation}
\label{eq:heatr2c}
\frac{\Gamma_c}{H n_H k_B} = \frac{h\Delta\nu_D}{k_B}
\frac{J_{c0}}{\tilde J_0}\, \int dx \left[1-\frac{J_c(x)}{J_{c0}}\right]
\equiv \sqrt{2 T\, T_0}\, \frac{J_{c0}}{\tilde J_0}\, I_c ~,
\end{equation}
where we have defined
$I_c = \int dx [ 1- J_c(x)/J_{c0} ]$, and
$T_0 = (h\nu_{\alpha})^2/(k_B\, m_Hc^2) = 1.29\times 10^{-3}$ K.
For the example shown in Figure \ref{Jcon}, the integral is the shaded
area, $I_c=16.7$. It is clear from equation (\ref{eq:heatr2c}) that if
the number of photons in the background is not greater than one per
hydrogen atom, the heating rate due to \lya scattering is not very large
because $T_0 \ll T$. In the example of $T=2.65$ K at $z=10$,
equation (\ref{eq:heatr2c}) yields 
$\Gamma_c/(H n_H k_B) = 1.4\, {\rm K}\, (J_{c0}/\tilde J_0)$.

  Similarly, for the injected photons with a $\delta$-function profile,
the heating rate is 
\begin{equation}
\frac{\Gamma_i}{H n_H k_B} = \frac{h\Delta\nu_D}{k_B}\,
\frac{J_{i0}}{\tilde J_0}\,
\left\{ \int_{-\infty}^0 dx \left[1-\frac{J_i(x)}{J_{i0}}\right] 
-\int_0^{+\infty} dx \frac{J_i(x)}{J_{i0}} \right\}
\equiv \sqrt{2T\, T_0}\, \frac{J_{i0}}{ \tilde J_0}\, I_i ~.
\end{equation}
We note that, for the more exact case in which photons are injected
with a Voigt profile in frequency, the heating rate is obtained by
subtracting from $J_i(x)$ the steady-state solution of the background
with no scattering (shown in Fig. 3 of RD). However, one can easily
prove that the result is exactly the same as what is obtained by
subtracting instead the step function spectrum that results from a
$\delta$ function of the injected photons.
%For more general case (e.g. a Voigt injection profile), 
%the heating rate is given by integrating the
%difference between the no-scattering (only redshift) steady state
%solution and the scattering solution. 

  In Fig.~\ref{Jinj}, we show the spectrum $J_i$ for two different
temperatures: $T_k =2.65$K and $T=1$K, for $\gamma_S = 1.44 \times
10^{-6}$. In the $T_k = 2.65$K case, we have $\Gamma_i/(H n_H k_B)=
-0.104 {\rm K} (J_{i0}/\tilde J_0) $. This negative heating rate means
that the scattering of these injected photons actually cools the gas,
because some of the photons are scattered to the blue side of the
spectrum and gain energy from the hydrogen gas. At lower gas
temperatures, as in the $T_k=1 $K case shown in the figure,
$\Gamma_i/(H n_H k_B)= 0.40\,  {\rm K}\, (J_{i0}/\tilde J_0) $, the
photons heat the gas. However, at $z=10$ the 
temperature of adiabatically expanding gas is 2.65 K, and there is no
known mechanism which could cool the gas below this temperature,
so the net effect of the injected photons must be to cool the gas.
In any case, both the heating rate due to continuum photons and the
cooling rate due to injected photons are very small.

\begin{figure}
\plottwo{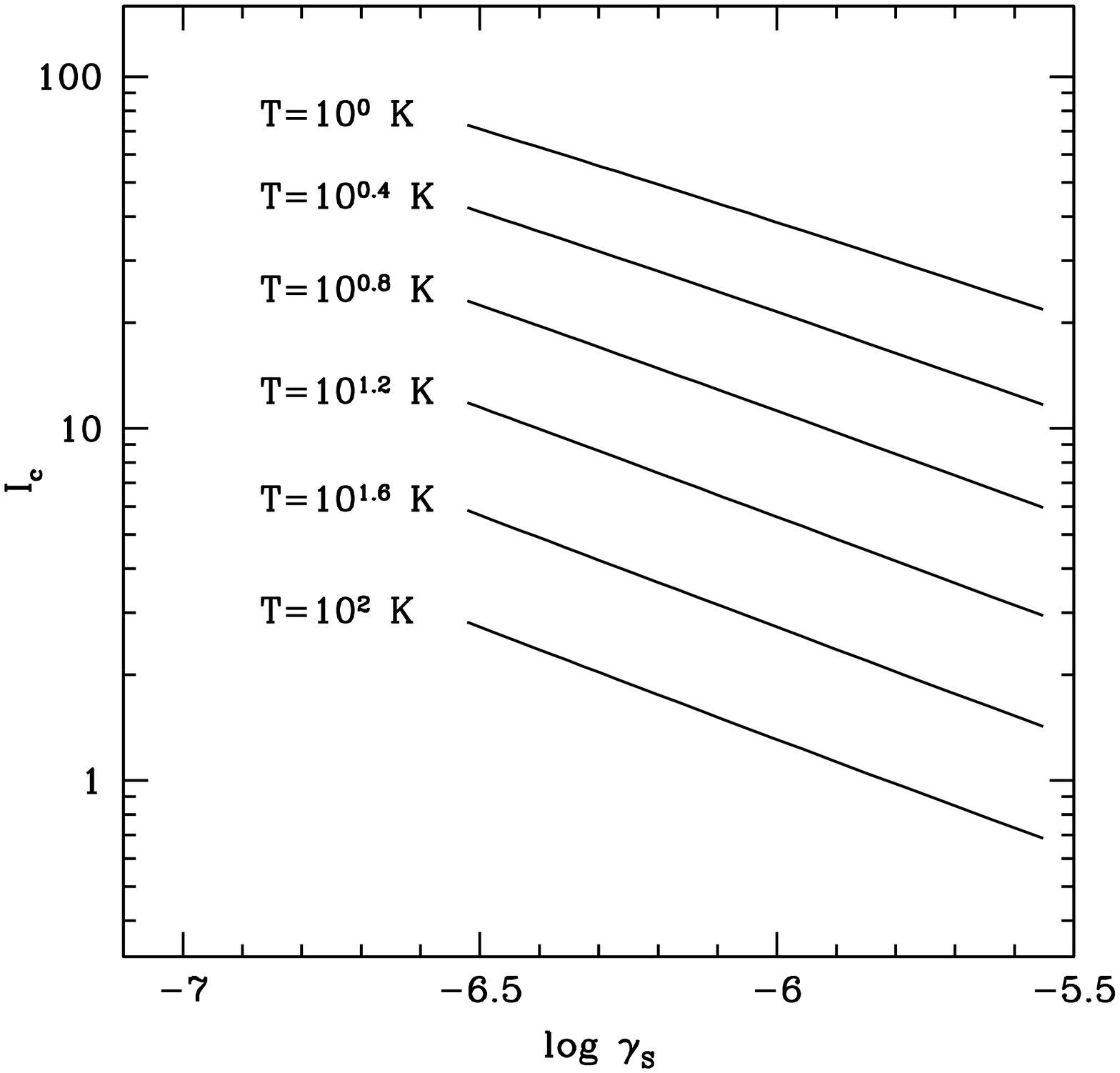}{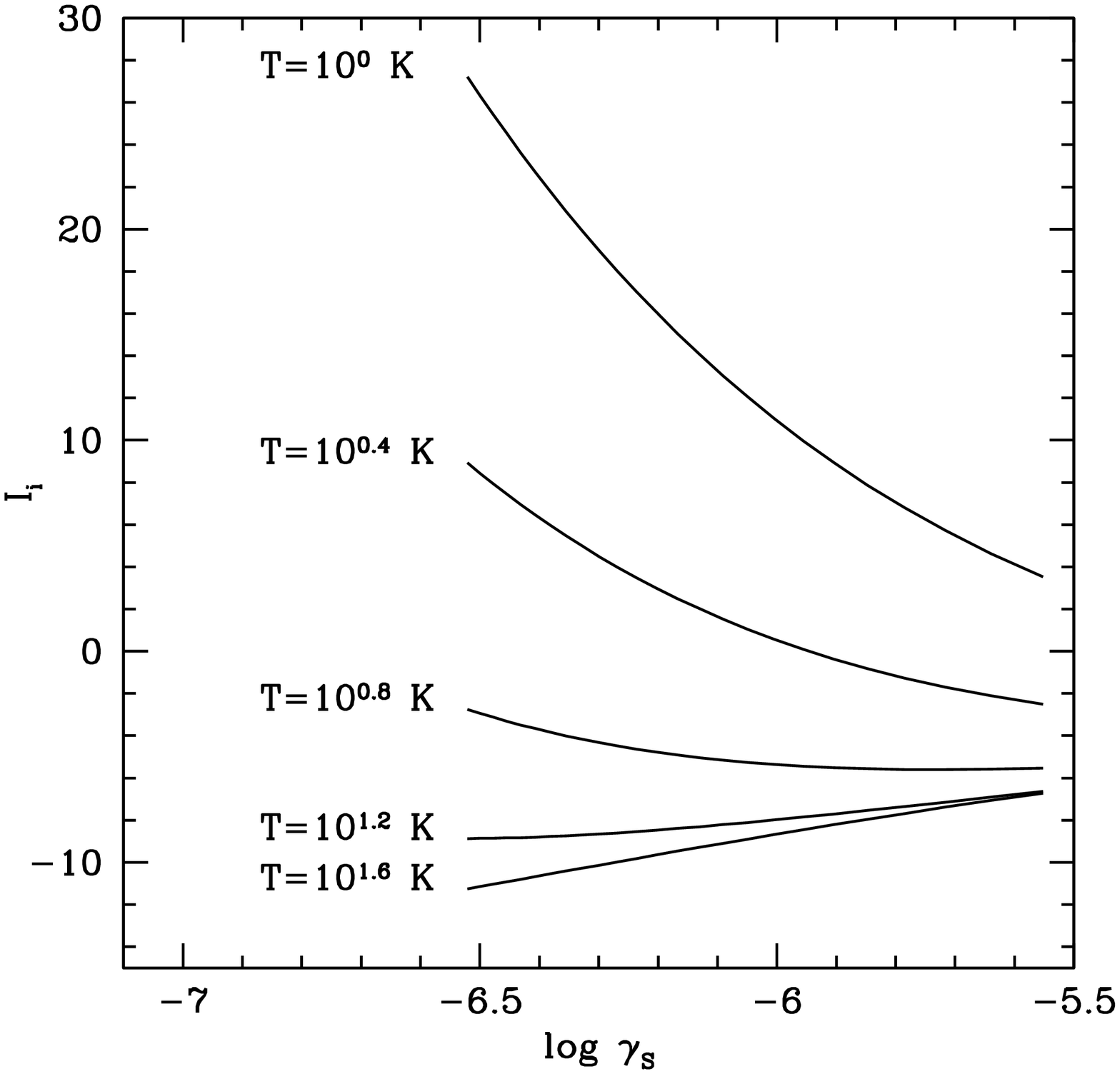}
\caption{\label{IcIi} Heating rate integral $I_c$ (left) and $I_i$
(right) versus $\log \gamma_S$ for
different temperatures.}
\end{figure}

\begin{figure}
\plottwo{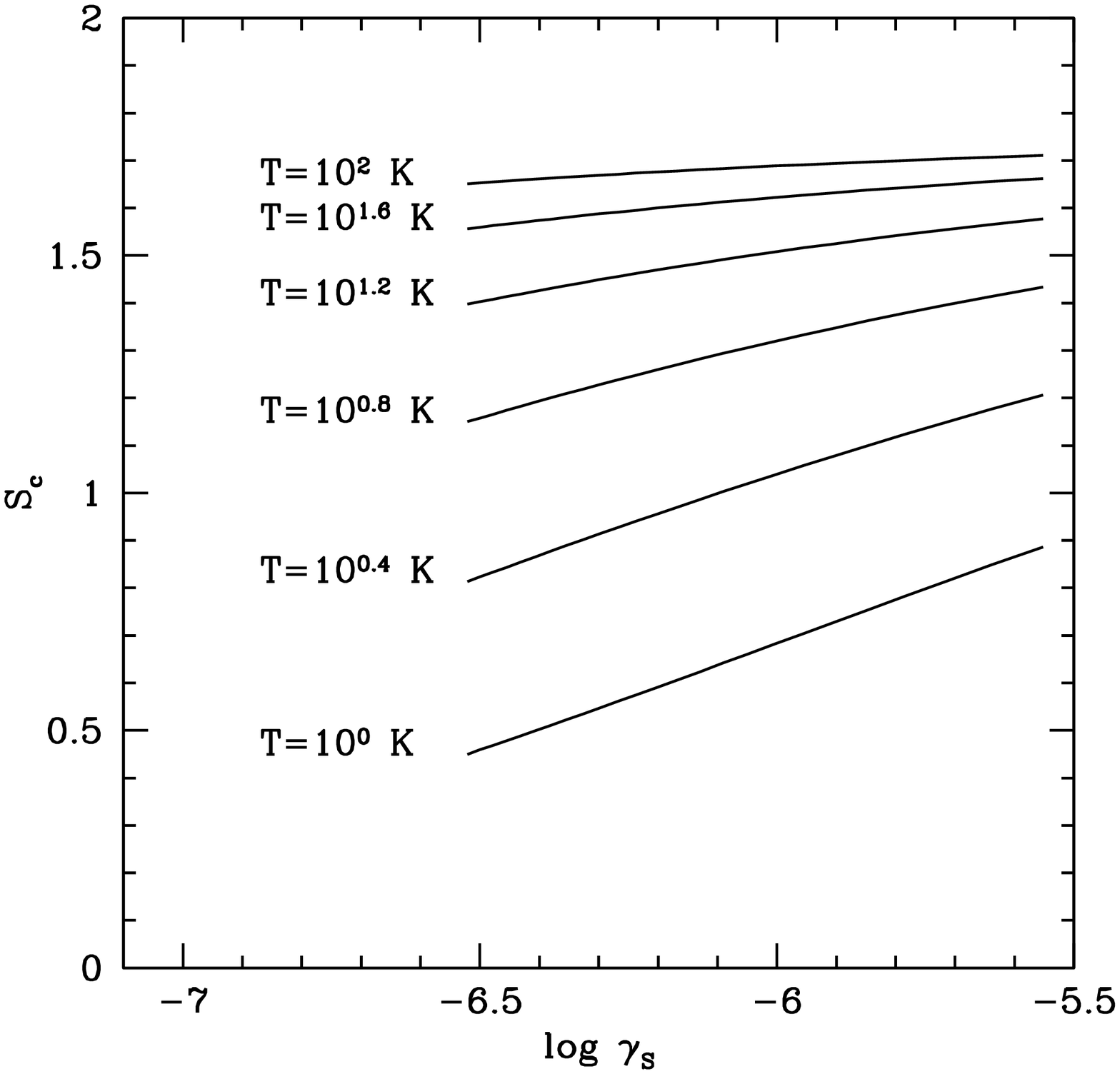}{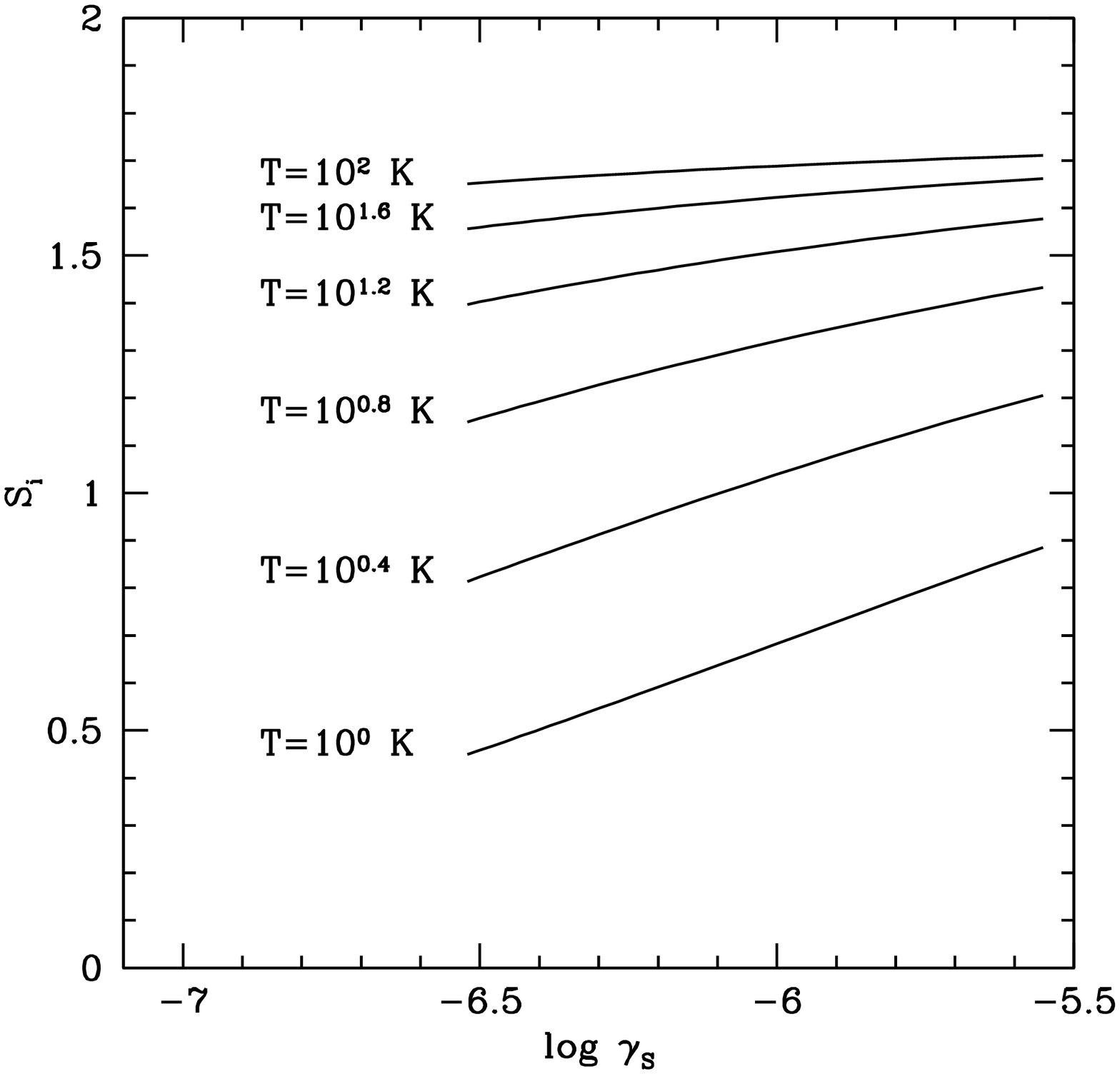}
\caption{\label{ScSi} Scattering rate integral $S_c$ (left) and $S_i$ (right)
versus $\log \gamma_S$ for
different temperatures.}
\end{figure}

  To calculate the spin temperature from equation (2), we also need to
know the scattering rate of the \Lya photons in equation
(\ref{eq:scatrate1}), which can be reexpressed as
\begin{equation}
P_{\alpha} = H \tau_{GP}
\int { J(x) \over \tilde J_0}\, \phi(x) \, dx ~.
\end{equation}
We define 
\begin{equation}
\label{eq:scati}
S_c \equiv \int \frac{J_c(x)}{J_{c0}} \phi(x)\, dx, \qquad
S_i \equiv \int \frac{J_i(x)}{J_{i0}} \phi(x)\, dx.
\end{equation}
We plot the integrals $I_c$, $I_i$, $S_c$, and $S_i$ as a function of
$\gamma_S = \tau_{GP}^{-1}$, for several different temperatures, in
Figures \ref{IcIi}-~\ref{ScSi}. Our code calculates the values of these
integrals for any given gas temperature and optical depth.

  The intensity corresponding to the thermalization rate required to
bring the spin temperature down to the kinetic temperature (see the
introduction) is
\begin{equation}
 {J_0\over \bar J_0} = {P_{\alpha}\over H\tau_{GP} S} =
{27 A_{10} T_{CMB} \gamma_S \over 4 T_{*}\, H S } ~,
\end{equation}
where the subscripts $c$ or $i$ can be applied. For the case
$\gamma_S=1.44\times 10^{-6}$ and $T=2.65$ K at $z=10$, $S_i$ or $S_c$
are close to one and we find $J_0/\bar J_0 \simeq 0.3$. Thus, from
our earlier result that the heating rate in this case is
$\Gamma_c/(H n_H k_B) = 1.4\, {\rm K}\, (J_{c0}/\tilde J_0)$ for the
case of continuum photons, we find that over the time that continuum
\lya photons bring the spin temperature down to the gas temperature,
they should heat the gas by only $\sim 0.5$ K, and injected photons
should cool the gas by $\sim 0.03$ K.

  The reason why the heating rate is much less than the value obtained
by assuming that each atom loses the same average energy at each
scattering as if it were at rest is that the intensity of the background
spectrum drops steeply at the \lya line center, as shown in Figures 1
and 2. The vast majority of the scatterings occur within a few thermal
widths of the line center, where the color temperature of the spectrum
(reflecting its slope) is almost equal to the kinetic temperature. When
these two temperatures are the same, there obviously cannot be any heat
exchange between photons and atoms because they are in thermodynamic
equilibrium. The heating from continuum photons occurs because the
color temperature is slightly larger than the kinetic temperature.
Injected photons have a temperature slightly lower than the kinetic
temperature, so they cool the gas. Cooling is caused by photons that
are on the red side of the line center because these photons are more
likely to be scattered by atoms moving in the direction opposite to
the photon (for which the Doppler effect can shift the photon frequency
to the line center), and these atoms will be slowed down by the
scattering. When the color temperature is below the kinetic temperature,
the cooling effect due to the preferential scattering of photons moving
in the opposite direction than the atoms is greater than the heating
effect due to recoil from photons moving in a random direction.

  The heating or cooling rates per unit volume are not proportional to
the atom density. If the atom density is increased, not only the number
of scatterings increases, but also the color temperature is brought
closer to the kinetic temperature because of the increased rate of
interaction between atoms and photons, and the average energy exchanged
between them at each scattering is reduced. The heating rate changes
only with the width of the background spectral feature that determines
the integrals $I_c$ and $I_i$, according to the energy conservation
argument used in \S 2.

  In the calculation of the spin temperature evolution from equation
(2), we assume that the color temperature of the \lya photons is equal
to the gas kinetic temperature. This is true at the line center in the
limit of large optical depth (Field 1958, 1959), but the slope of the
photon distribution varies outside the line center. In the more general
case, equations (\ref{eq:Ts})-(\ref{eq:y}) can be generalized as 
\begin{equation}
y_{\alpha} T_k \to \int dx ~ y_{\alpha} (x) T_{\alpha}(x) =
\frac{T_*}{A_{10}} \frac{4}{27} \int d P_{\alpha}(x)
\end{equation}
\begin{equation}
y_{\alpha} \to \int dx ~ y_{\alpha}(x) = \frac{T_*}{A_{10}}\frac{4}{27}
\int d P_{\alpha}(x)/T_{\alpha}(x) 
\end{equation}
where the color temperature $T_{\alpha}$ at $x$ is defined by 
\begin{equation}
\frac{J'(x)}{J(x)} \equiv -
\frac{h\nu}{k_B T_{\alpha}(x)} \approx -\frac{h\nu_0}{k_B T_{\alpha}(x)},
\end{equation}
and
\begin{equation}
d P_{\alpha}(x)= 4\pi\sigma_0\, \Delta\nu_D\, J(x) \phi(x) dx
\end{equation}
Because the vast majority of scatterings occur near the line center,
where the color temperature and kinetic temperature are very nearly
the same, this generalization does not significantly change the result
for the spin temperature. We have found by a numerical test that this
correction affects the spin temperature by only a small fraction of a
percent.

\section{Simple Models for the Thermal Evolution}

  We will now consider a simple model where the emissivity of \lya
photons turns on at a redshift $z_i$ and increases linearly with
redshift thereafter. We will also consider the effect of X-ray emission
from the same star formation regions, which dominates the heating
rate of the gas. This will illustrate a plausible thermal history of
the gas and the expected strength of the absorption or emission signal
in the redshifted 21cm line.

  The kinetic temperature of the gas in the expanding Universe evolves as 
\begin{equation}
(1+z)\frac{d T_k}{d z}
=2 T_k-\frac{2}{3}\frac{(\Gamma_{tot}-\Lambda_{tot})}{(H n k_B)} ~,
\end{equation}
where $\Gamma_{tot}$ and $\Lambda_{tot}$ are the total heating and
cooling rates, respectively, and $n$ is the total gas number density.

  In the absence of heating and cooling, the gas temperature decreases
adiabatically  with $T \propto (1+z)^2$. We use the code RECFAST 
\footnote{\url{http://www-cfa.harvard.edu/$\sim$sasselov/rec}}
\citep{sss99,sss00}
to calculate the temperature evolution of the gas before the
first stars and quasars turn on. We note that at $z\gtrsim 40$ the spin
temperature drops below the CMB temperature due to the collisional
coupling (in Fig. 6, this drop is seen at the highest redshift end). The
corresponding absorption is, however, difficult to observe at the very
long wavelengths corresponding to this redshift. At lower redshifts, the
spin temperature is practically equal to the CMB temperature until the
first \lya photons are produced.

\subsection{\lya emissivity}

  If the comoving photon emissivity (defined as the number of photons
emitted per unit comoving volume, time and frequency) at the \lya line
from stars at redshift $z$ is $\epsilon(\nu_{\alpha}, z)$, then the
{\it comoving} \lya photon intensity at redshift $z$ is given by
\begin{equation}
J_{c0}(z) = \frac{1}{4\pi}(1+z)^{\alpha_S} \int_z^{z_{max}}
\frac{c}{H(z')} \epsilon (\nu_{\alpha}, z') (1+z')^{-(1+\alpha_S)} dz',
\end{equation}
where the emissivity spectrum is assumed to be $\epsilon(\nu) \propto
\nu^{-\alpha_S - 1}$, and $z_{max}$ is determined by the condition that
only the photons emitted at frequencies lower than the Ly$\beta$ line
will not interact with the intergalactic medium:
\begin{equation}
(1+z_{max})=\frac{\nu_{\beta}}{\nu_{\alpha}}(1+z)= \frac{32}{27}(1+z)
\end{equation}

  To discuss the plausible range of values of the comoving emissivity,
it is convenient to introduce the unit
\begin{equation}
\tilde \epsilon \equiv {n_H\, H\over \nu_{\alpha} } =
{4\pi \over c}\, H \tilde J_0 ~.
\end{equation}
Thus, if $\epsilon = \tilde \epsilon$, one photon is being emitted
per hydrogen atom and per Hubble time, per unit $\log(\nu)$.

  During the epoch of reionization, at least one photon must be emitted
for every baryon in the universe. For a homogeneous medium, the
recombination rate $\alpha n_e^2$ is equal to the Hubble rate, $H$,
at $z\simeq 6$; hence, if reionization takes place during the interval
$20 \gtrsim z \gtrsim 6$, a few photons should be emitted for every
baryon to ionize them and to compensate for recombination. The number
of recombinations can be increased due to the clumping factor of the
ionized intergalactic gas (see, however, \citealp{MHR00}
 for a discussion of why the clumping factor of ionized gas is
probably not much larger than unity during the epoch of reionization).
If these ionizing photons are produced by star-forming regions, then
the photon emissivity per unit $\log(\nu)$ between \lya and the Lyman
limit should be 3 to 5 times larger than the emissivity of ionizing
photons, owing to the average Lyman break of the atmospheres of massive
stars with a Salpeter Initial Mass Function (e.g., \citealp{bc93}). 
The photons between \lya and the Lyman limit may have a larger
escape fraction from the star formation regions than the ionizing
photons (if some of the absorption is due to local hydrogen with low
dust content), in which case the ratio of the emissivity between \lya
and Lyman limit to the ionizing emissivity would be further increased.
This suggests that the emissivity near \lya should be at least
$\epsilon \gtrsim 10 \tilde \epsilon$.

  The total number of ionizing photons emitted is also related to the
metallicity produced by massive stars. To produce one ionizing photon
per baryon, the mean metallicity of the universe must be raised by
$10^{-5}$, or $5\times 10^{-4} Z_{\odot}$ (e.g., \citealp{MS96}).
If the typical metallicity in the \lya forest and damped \lya systems
at $z\gtrsim 3$, which is $Z\sim 10^{-2} Z_{\odot}$, is not to be
exceeded, one cannot produce more than $\sim 20$ ionizing photons per
baryon, implying that $\epsilon \lesssim 100 \tilde \epsilon$ at \lya.

\begin{figure}
\plotone{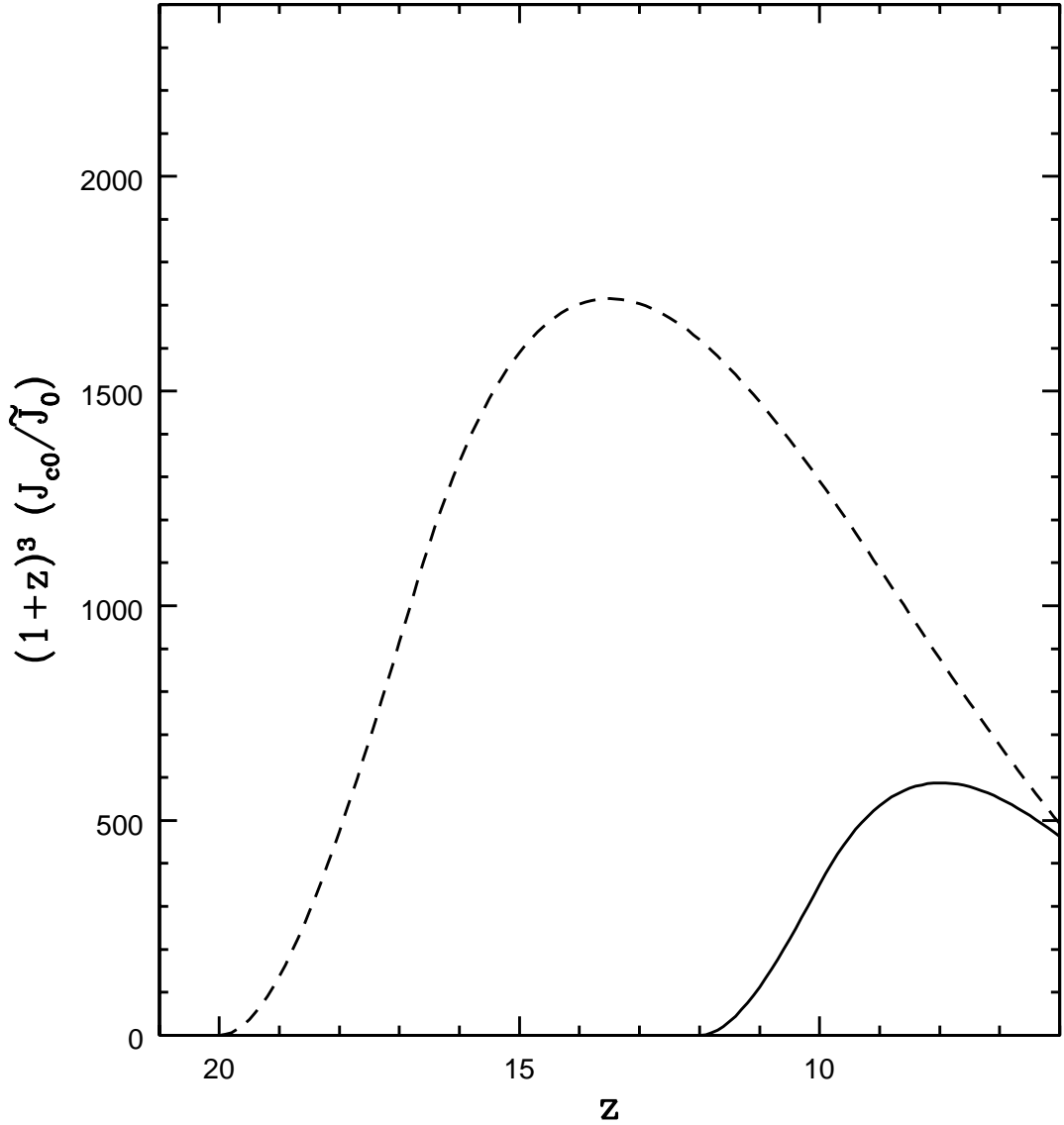}
\caption{\label{fig:jeps} The Lyman alpha proper intensity, plotted as
$(1+z)^3 (J_{c0}/\tilde J_0)$, as a function of redshift. Dashed and
solid curves are for $z_i=20$ and $z_i=12$, respectively. We also assume
$J_{i0}=J_{c0}$}
\end{figure}

  As a simple model of the \lya background evolution, we assume that
the \lya emissivity increases linearly from 0 at some initial 
redshift $z_i$,
\begin{equation}
\label{eq:epsz}
\epsilon(z) = \left\{ 
\begin{array}{ll}
0; &z > z_i\\
\epsilon_r \frac{z_i -z}{z_i-z_r}; & z_r < z < z_i\\
\end{array}\right.
\end{equation} 
We will use two values for the peak emissivity at \lya,
$\epsilon_r$, reached at $z_r$: $\epsilon_r=10 \tilde \epsilon$ and
$\epsilon_r = 100 \tilde \epsilon$, based on the previous discussion.
Assuming also a flat spectrum,
$\alpha_S=0$, the \lya intensity due to the continuum photons is then 
\begin{equation}
{ J_{c0}(z)\over \tilde J_0} =  {\epsilon_r \over \tilde \epsilon} \,
 g(1+z) ~,
\end{equation}
where the function $g$ is
\begin{equation}
g(x) = {x^{3/2} \over z_i - z_r }\,
\left[\frac{2}{3} (1+z_i)\left(x^{-3/2}-x_<^{-3/2}\right)
-2\left(x^{-1/2}-x_<^{-1/2}\right)\right] , \qquad z<z_i
\end{equation}
and 
\begin{equation}
x_< = \min\left[(1+z_i), \left(\frac{32}{27}x-1\right)\right].
\end{equation}
The resulting proper \lya intensity is plotted as the function
$(1+z)^3 J_{c0}/\tilde J_0$ in Fig.~\ref{fig:jeps}, for $z_r=6$ and
$\epsilon_r/\tilde\epsilon=10$, for two cases of initial redshift:
$z_i=20$ and $z_i=12$. The proper \lya intensity increases rapidly
after $z_i$ and then decreases as $z$ decreases due to the $(1+z)^3$ factor.

  We will assume that $J_{i0} = J_{c0}$. The injected 
photons have a comparable intensity to the continuum 
ones, because the photons emitted between
Ly$\beta$ and the Lyman limit are similar to those emitted between
\lya and Ly$\beta$, and a large fraction of the Lyman continuum photons
will eventually result in an injected \lya photon when the atom that
they ionize recombines again.

\subsection{Heating by soft X-rays}

  Since the scattering of \Lya photons produces a negligible heating
rate for the gas, the main heating mechanism should be soft X-ray
photons. The soft X-rays can be emitted by a number of sources
associated with star-forming regions, such as X-ray binaries, supernova
remnants and QSOs.

  X-ray photons have a photoionization cross section
that is much smaller than that of the ionizing photons near the hydrogen
ionization potential, so they can penetrate deep into the neutral
regions of the intergalactic medium. There, they photoionize hydrogen
or helium, and the high-energy electron that is produced can then
transfer a fraction of its energy to the gas by Coulomb collisions
with other electrons and protons. Once the energy has been thermalized
among the charged particles at a temperature low enough to avoid losses
by collisional excitation of hydrogen, the thermal energy can be
transferred to the hydrogen atoms in slow collisions involving dipole
interaction. The fraction of the X-ray energy that can be converted to
the gas thermal energy (rather than being used in collisional
ionizations and excitations of hydrogen or helium) depends on the
fractional ionization of the gas. Using the RECFAST code, we found
that the initial ionized fraction from the recombination 
era (before more ionizations
are caused by the X-rays) is $2\times 10^{-4}$, which implies that a
fraction of 14\% of the X-ray energy is used to heat the gas
\citep{SvS85}. As the fraction of free electrons increases, this
fraction also increases, but for simplicity we shall assume it a
constant.

  We parameterize the emissivity in X-rays in terms of the fraction
of the energy that is emitted in X-rays compared to the energy emitted
at \lya per unit $\log\nu$, which we designate as $\alpha_x$. The X-ray
heating rate is then
\begin{equation}
\Gamma_x(z) = 1.4 \times 10^{-3} \left(\frac{\alpha_x}{0.01}\right)
h\, \epsilon(z) ~,
\end{equation} 
where $\epsilon(z)$ is given in Eq.~(\ref{eq:epsz}), and $h$ is the
Planck constant. 
At $z=6$, the heating rate due to X-rays in this model is 
\begin{equation}
\frac{\Gamma_x}{n_H H k_B} = 
1.7 \times 10^3 \,\left(\frac{\alpha_x}{0.01}\right)\,
{\epsilon_r \over 10\tilde\epsilon} \,{\rm K} ~.
\end{equation}
Therefore, by emitting only $\sim$ 1\% of the energy as X-rays, the
sources can rapidly heat the intergalactic medium to temperatures
above the CMB.

  \subsection{Results for thermal evolution and the 21 cm Antenna
Temperature}

\begin{figure}
\plottwo{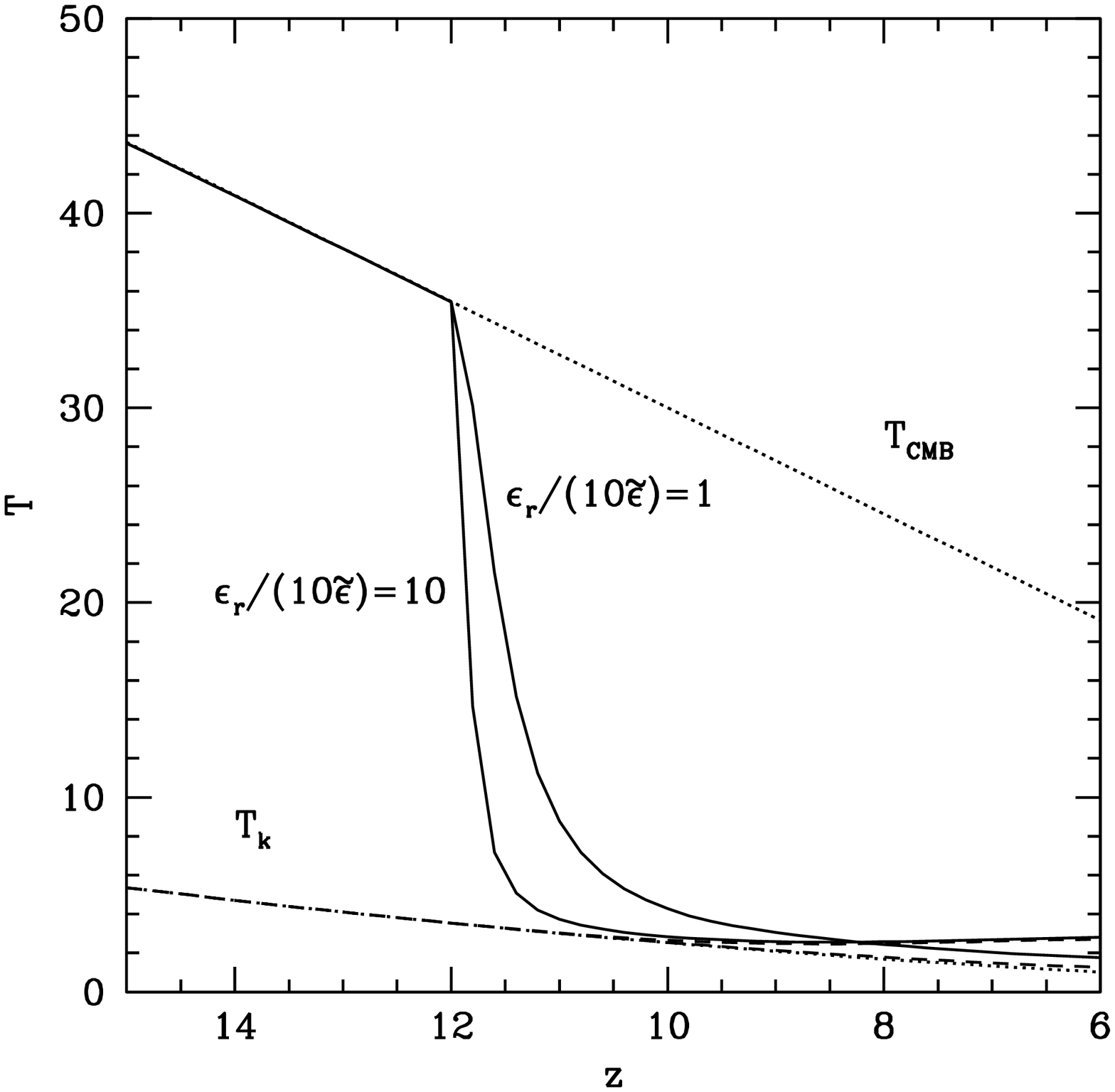}{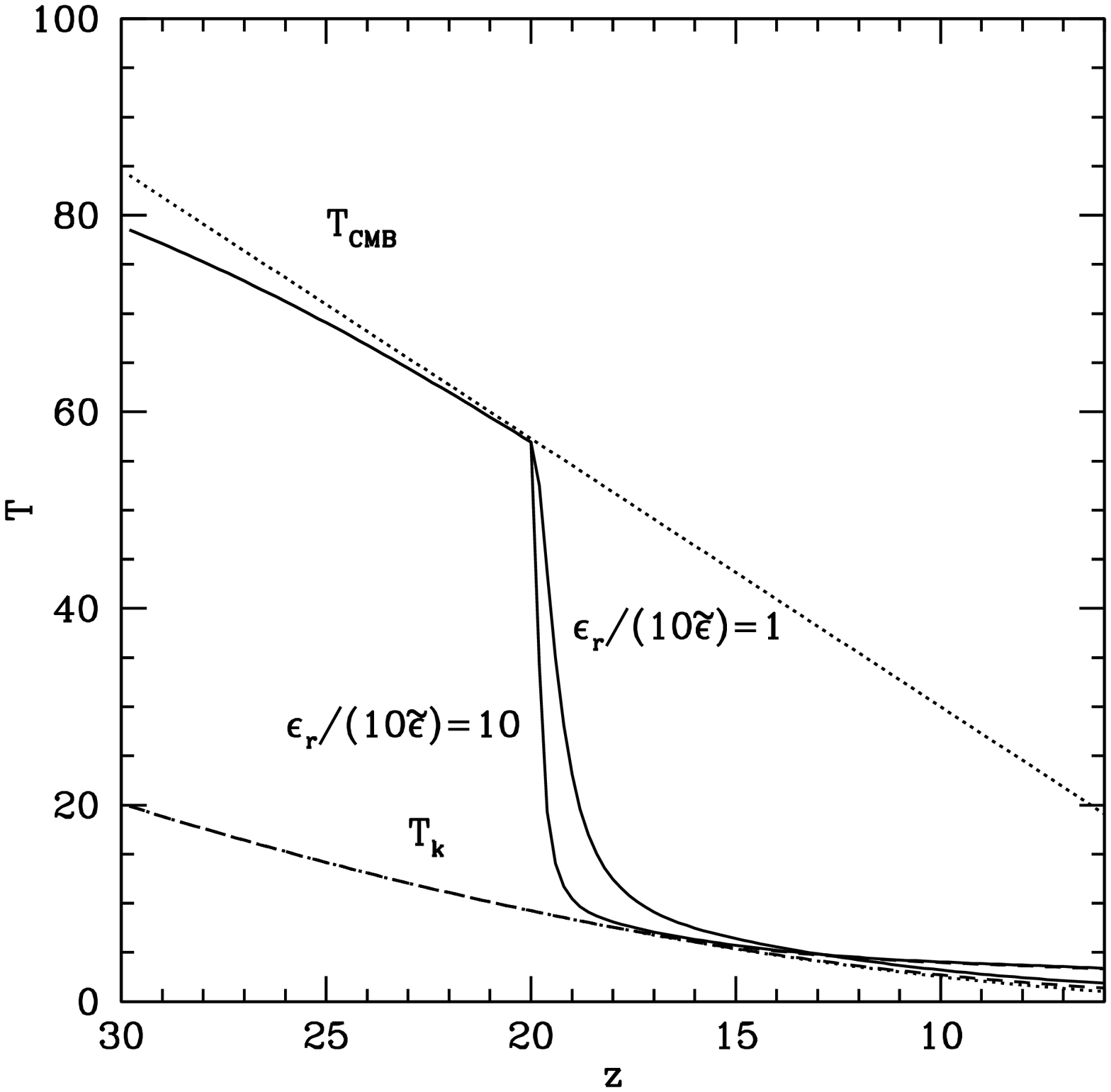}
\caption{\label{fig:Ts} The kinetic and spin temperature evolution of the gas with 
only \Lya photons which increases linearly after some initial redshift
$z_i$. The CMB temperature and adiabatic evolution of gas kinetic
temperature is plotted in dotted lines, \lya heated kinetic
temperature in dashed lines, and spin temperature in solid lines. The
two curves are for fudge factor $\frac{\epsilon_r}{10 \tilde
  \epsilon}=1$ and 10.
{\bf Left:} The ``late start'' scenario with
$z_i=12$. {\bf Right:} The ``early start'' scenario with
$z_i=20$. Results for two \lya emissivity models are plotted in each
figure, with fudge factor $\frac{\epsilon_r}{10 \tilde  \epsilon}=1$ and 10. 
}
\end{figure}

  The evolution of the gas kinetic and spin temperatures in the absence
of X-ray heating is plotted in Fig.~\ref{fig:Ts}, for two values of the
initial redshift ($z_i=20$ and $z_i = 12$), and two values of the
emissivity ($\epsilon_r/\tilde\epsilon = 10 $ and $100$).

  In the absence of X-rays, the heating due to \lya scattering is very
low and the kinetic temperature of the gas deviates only slightly from
the adiabatic curve at late times. For our assumed emissivity of 10 to
100 photons per baryon per Hubble time, the spin temperature drops to
values very close to the kinetic temperature, producing a large
absorption signal on the CMB of up to $200$ mK, as seen in Figure 7.

  Figure 8 shows the thermal evolution and the resulting antenna
temperature when we include X-ray heating with $\alpha_x = 0.01$,
for the case $z_i=12$ and $\epsilon_r/\tilde\epsilon = 10$. With
only 1\% of the energy emitted as X-rays, the kinetic temperature of
the gas increases rapidly when the first sources turn on, leaving
only a brief period of absorption in the beginning of small amplitude,
which then turns to emission, with an amplitude of less than 20 mK.

\section{Conclusion}

  We have calculated the \Lya photon scattering and heating rates in
the atomic intergalactic medium, including both redshifted continuum
photons emitted between \lya and Ly$\beta$, and injected photons produced
by recombinations or excitations by Ly$\beta$ and higher Lyman series
photons. The heating rate of the gas due to the scattering of these \Lya
photons is very small and essentially negligible. In the absence of
other heating sources, the gas would continue to cool down almost
adiabatically while the spin temperature would rapidly drop to nearly
the kinetic temperature given a reasonable emission of photons during
the reionization epoch. This would produce a 21cm absorption against
the CMB as large as $\sim$ 200 mK, at the average hydrogen density of
the universe.

  Of course, if the absorption signal can be observed at
high angular and frequency resolution, the antenna temperature should
fluctuate around its mean value owing to density fluctuations, as well
as spin temperature fluctuations that would be produced by spatial
variations in the history of the photon emissivity between \lya and
the Lyman limit (e.g.\citealp{TMMR00,CM03}). 
At the same time, during the epoch of reionization,
there should be a gradual increase of the fraction of the volume that
is ionized (e.g., \citealp{M02}), and the average antenna
temperature should then be multiplied by the fraction of baryons that
remain in atomic form. Because this fraction of baryons should
decrease gradually, any 21cm absorption or emission should also vary
smoothly with redshift.

  Unfortunately, the large absorption signal predicted in the absence
of X-ray heating is unlikely to be present. It is very likely that
significant X-ray emission is produced within only a few million years
of the formation of the first stars in the universe: some of the stars
may be in binaries which will evolve into high-mass X-ray binaries,
and supernova explosions can also emit soft X-rays (see e.g. 
\citealp{O01,VGS01}). Photons with
energies as low as $0.2$ keV can propagate as far as 3 comoving Mpc
over the atomic intergalactic medium at $z\sim 10$ before they are
absorbed by hydrogen or helium.

  There may still be some ways in which certain regions of the
intergalactic medium may be illuminated by \lya photons from the first
stars, without receiving soft X-rays as well. For example, if stars form
in a region with a high column density of hydrogen and helium with very
little dust, and the hydrogen and/or helium is not fully ionized for a
long time, the photons between \lya and the Lyman limit from stars would
escape, while the ionizing and soft X-ray photons might all be locally
absorbed. In particular, there is the possibility that singly ionized
helium remains self-shielded as massive stars ionize the gas around
them. We note from Figure 6 that the time required to bring down the
spin temperature can be very short ($10^8$ years or less) for typical
values of the \lya background intensity, and it is possible that a
complete absorption of soft X-rays from the first star-forming regions
would persist long enough to allow for a strong 21cm absorption signal
in local regions of the intergalactic medium, when they are illuminated
by the first light from stars forming in their vicinity. Nevertheless,
it seems likely that any decrease of the spin temperature will be
short-lived, and that X-ray heating will cause the 21cm signal to switch
to emission a short time after the first stars are formed.

  Future observations of the 21cm signal in absorption and emission
at high angular and frequency resolution have an enormous potential
for revealing the details of the reionization epoch. Fluctuations in
the antenna temperature arise from both gas density and spin
temperature variations. The density fluctuations will inform us of the
state of gravitational evolution of primordial fluctuations at the
highest attainable redshifts after recombination, while the spin
temperature fluctuations will reveal the spatial distribution of the
first sources of ultraviolet and X-ray emission in the universe.

\acknowledgments

   This work was supported in part by NSF grants NSF-0098515
and PHY99-07949. 

\appendix
\section*{Appendix}

  MMR assumed that the heating rate per unit volume of the hydrogen gas
due to \lya scattering is equal to the recoil energy given to a
hydrogen atom at rest, $(h\nu_{\alpha})^2/(m_Hc^2)$, times the
scattering rate:
\begin{equation}
\Gamma_{MMR} = {(h\nu_{\alpha})^2\over m_Hc^2}\, n_H\, P_{\alpha} =
{(h\nu_{\alpha})^2\over m_Hc^2}\, 4\pi n_H\, \sigma_0\, \Delta\nu_D\,
\int J(x)\phi(x)\, dx ~.
\end{equation}
This heating rate was justified from equation (\ref{eq:FP}) for the time
evolution of the photon intensity $J$. MMR assumed that the rate of heat
transfer to the atoms is determined by the change in the spectrum due to
the recoil effect, which is the second term in the right-hand-side of
equation (\ref{eq:FP}):
\begin{equation}
\left( {\partial J\over \partial \tau} \right)_{recoil} = 
{\partial (\eta\phi J) \over \partial x} ~.
\end{equation}
They then computed the heating rate by integrating the variation in
photon energy density per unit physical time over frequency:
\begin{equation}
\Gamma_{MMR} = 4\pi h \Delta\nu_D\, n_H\sigma_0 
\int dx \, x {\partial (\eta\phi J)\over\partial x} = (H n_H k_B)
T_0 \tau_{GP}\, {J_{c0}\over \tilde J_0}\, S_c ~,
\end{equation}
where the integral $S_c$ is given in equation (\ref{eq:scati}), and
$\tau_{GP}$ is the Gunn-Peterson optical depth (all symbols are
defined in \S 2 and 3; the last equality involves an integration by
parts).

  However, the first term in equation (\ref{eq:FP}) must also be
included to compute the heating rate of the gas. In fact, the heating
rate of the gas is not just determined by the recoil, but also by the
fact that the background intensity is greater on the red wing of the
\lya line than on the blue wing, and scatterings have an average
tendency to bring photons back into the line center. The energy change
of the photons due to the first term in equation (\ref{eq:FP}) can only
go into the gas (whereas that of the third term is due to redshift
and goes into the expansion of the universe, and that of the fourth
term is provided by a source). Thus, the correct heating rate is
\begin{equation}
\Gamma = 4\pi h \Delta\nu_D\, n_H\sigma_0 
\int dx \, x {\partial \over\partial x} 
\left( {\phi\over 2} {\partial J \over \partial x} + \eta\phi J
\right) ~.
\end{equation}
Under steady-state conditions (and no source term; here, we consider
continuum photons only, to simplify the discussion), we can replace the
quantity inside parenthesis by $- \tau_{GP}^{-1} \partial J/\partial x$,
and integrating by parts, we obtain
\begin{equation}
\Gamma = (H n_H k_B)\, \sqrt{T_0\, T}\, 
{J_{c0}\over \tilde J_0}\, I_c ~,
\end{equation}
which is the expression in equation (\ref{eq:heatr2c}). Note that,
because $\tilde J_0 = (n_H c)/(4\pi \nu_{\alpha})$, the heating rate
per unit volume is independent of the atomic density, except for the
weak dependence of $I_c$ on $\tau_{GP}$, whereas MMR assumed that
the heating rate {\it per atom} is independent of $n_H$.

\clearpage


\clearpage

%\begin{figure}
%\plottwo{T1.ps}{T2.ps}
%\caption{\label{T} {\bf Left}: The kinetic and spin temperature evolution of the
%gas with only \Lya photon and X-ray photon from halos. The gas kinetic 
%temperature is almost identical to adiabatic value until after z=8.
%The spin temperature drops below CMB temperature after \Lya is turned on.
%{\bf Right}: The same plot with additional X-ray heating, gas kinetic temperature
%raises above $T_{CMB}$. 
%}
%\end{figure}

%\begin{figure}
%\plottwo{dT1.ps}{dT2.ps}
%\caption{\label{dT} The antenna temperature difference as a function
%of redshift. {\bf Left}: only \Lya photon is present, a strong
%absorption feature is produced. {\bf Right}: With X-ray photon
%heating, the absorption feature is much shallower, and followed by
%emission.
%}
%\end{figure}

\begin{figure}
\plottwo{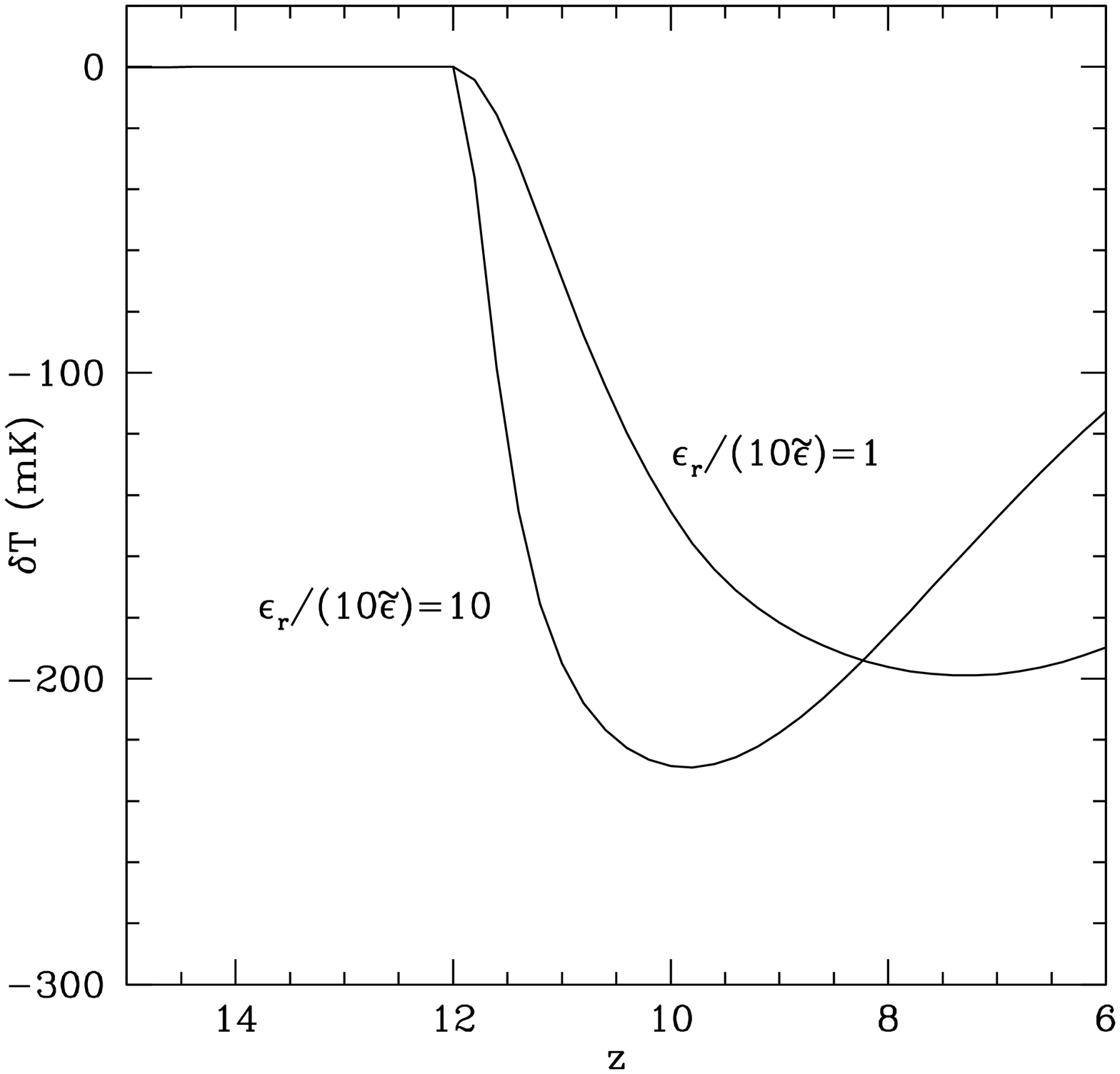}{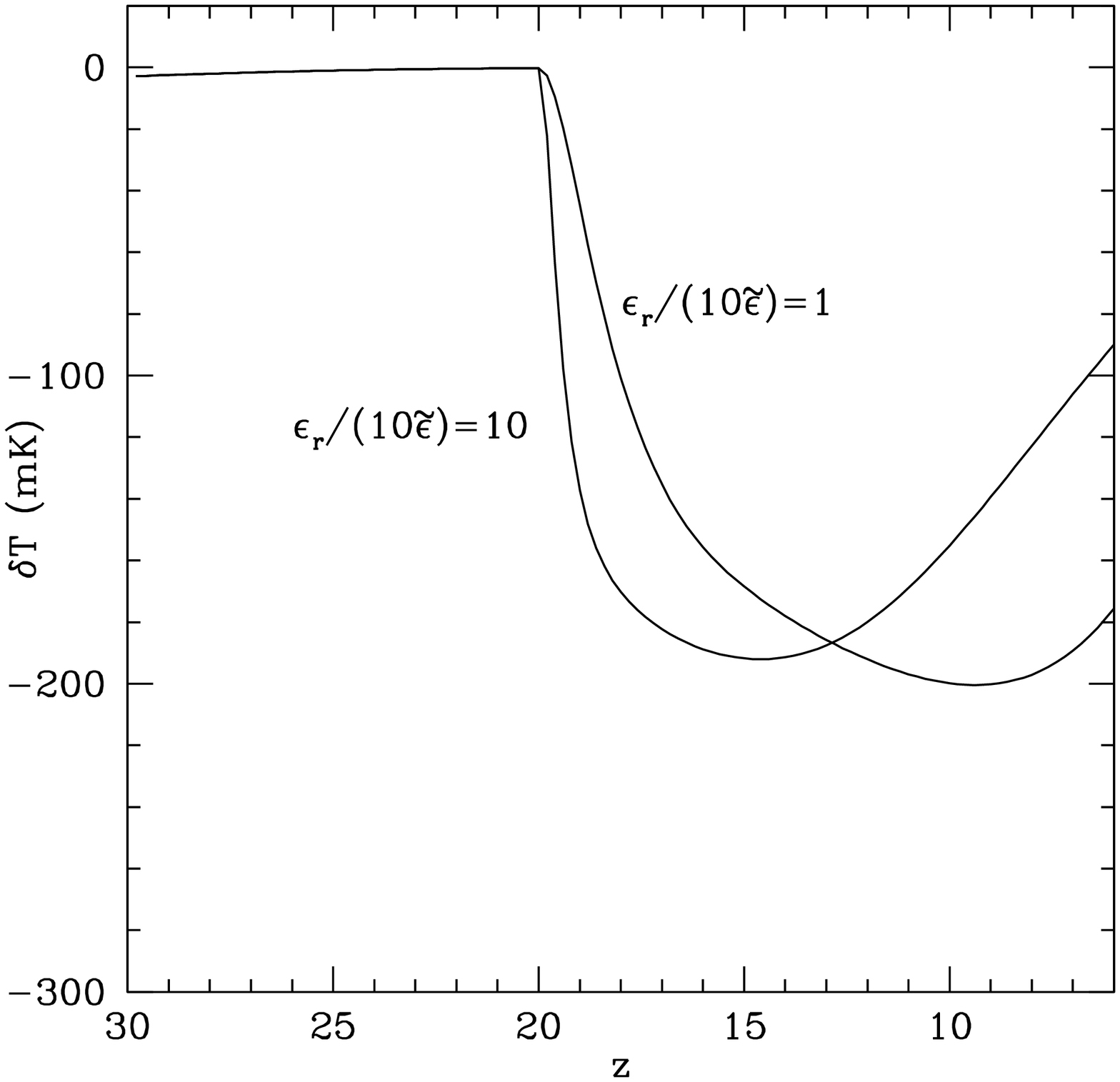}
\caption{\label{fig:dT} The antenna temperature difference as a function
of redshift with only \lya photons. {\bf Left}: late start $z_i=12$;
{\bf Right}: early start $z_i=20$.
}
\end{figure}

\begin{figure}
\plottwo{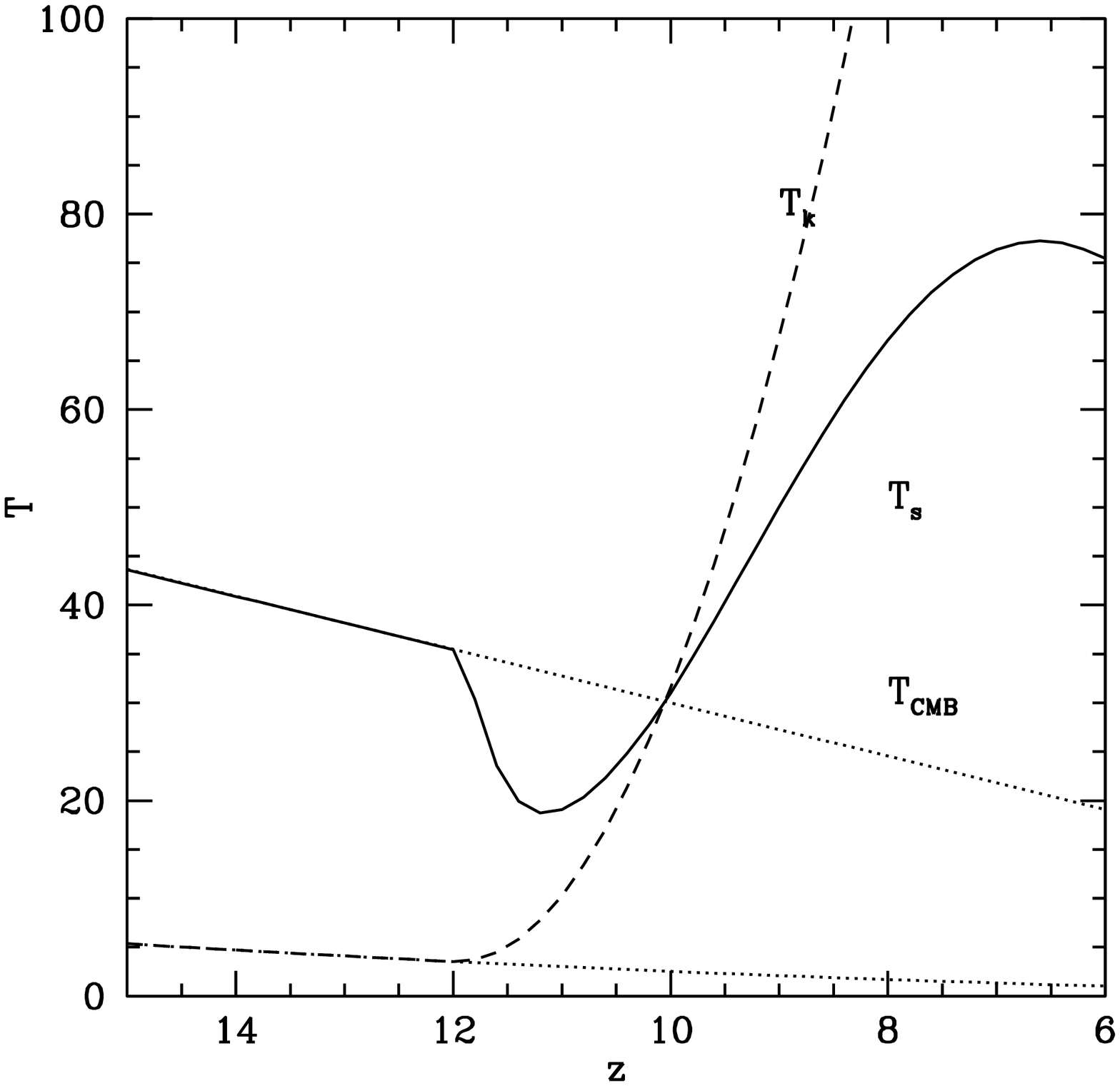}{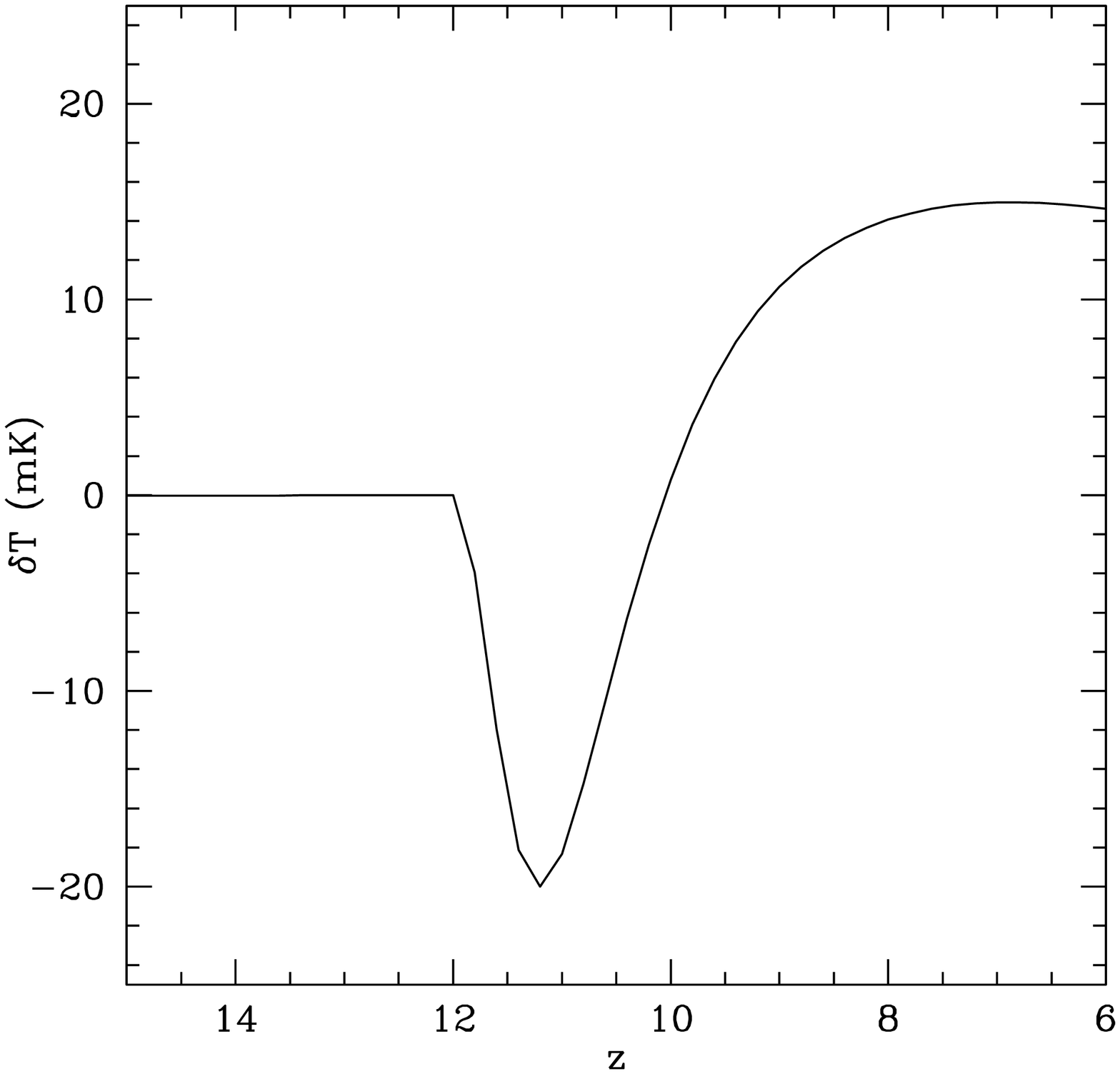}
\caption{\label{fig:Tx} Model with X-ray heating, $z_i=12$, 
$\frac{\epsilon_r}{10 \tilde  \epsilon}=1$, $\alpha_x=0.01$.
{\bf Left}: evolution of temperatures; 
{\bf Right}: the antenna temperature difference.}
\end{figure}

\clearpage 

%
%% Tables should be submitted one per page, so put a \clearpage before

%% Two options are available to the author for producing tables:  the
%% deluxetable environment provided by the AASTeX package or the LaTeX
%% table environment.  Use of deluxetable is preferred.
%%

%% Three table samples follow, two marked up in the deluxetable environment,
%% one marked up as a LaTeX table.

%% In this first example, note that the \tabletypesize{}
%% command has been used to reduce the font size of the table.
%% Note also that the \label command needs to be placed 

\clearpage

%% Tables may also be prepared as separate files. See the accompanying
%% sample file table.tex for an example of an external table file.
%% To include an external file in your main document, use the \input
%% command. Uncomment the line below to include table.tex in this
%% sample file. (Note that you will need to comment out the \documentclass,
%% \begin{document}, and \end{document} commands from table.tex if you want
%% to include it in this document.)


%% The following command ends your manuscript. LaTeX will ignore any text

\end{document}